# Extreme high-pressure distortion phase of $Bi_2MoO_6$ in $Bi_2Mo_{0.94}W_{0.06}O_6$ at ambient pressure for augmented octahedral rotations and metal-oxide charge transport

**Anurag Pritam[a)] and Vaibhav Shrivastava[a),*]**

a) Dielectric Laboratory, Shiv Nadar University, Uttar Pradesh, India

Bismuth molybdate (BMO) is the simplest compound of aurivillius family with a wide range of application as a tunable capacitor and sensor. The present article deals with the study of astructural, optical and catalytic property of tungsten doped Bismuth molybdate. The crystal structure is critically studied by Rietveld structural refinement method by using X-ray powder diffraction data. Raman scattering indicates that the structural distortion in BMO is mainly due to the rigid rotation of $MoO_6$ octahedra. Due to the introduction of tungsten, the system experienced an increase in the band gap, which is studied by Uv-vis spectroscopy and W doped BMO show remarkable photocatalytic behavior compared to pure one.

## 1. Introduction

Tailoring the crystallinity and bandgap of perovskite materials has phenomenological significance owing to their numerous applications in the field of optoelectronics, memory devices, sensing, and photocatalysis. Among all materials, Bismuth molybdates ($Bi_2MoO_6$) serves as an ideal material with all these unique properties for LTCC technology, it is one of the simplest units of the large family of Bismuth layered structured ferroelectrics(BLSF) perovskite known since 1949.[1-2] The structure of this compound is described by the general formula $(Bi_2O_2)^{2+} (A_{n-1}B_nO_{3n+1})^{2-}$, where A sites can be taken by mono, di or trivalent compound and B sites can be occupied by rare-earth elements and m represents the number of $BO_6$ octahedra in a perovskite layer (n=1,2,3,4 and 5).[3-5] The crystal arrangement of this compound comprises of alternating $Bi_2O_2$ layers and perovskite-like a layer of corner linked $BO_6$ octahedra.[3-5] $Bi_2MoO_6$ known as one of the most convoluted compounds of BLSF due to the polymorphic phase of its crystal structure, which can be isolated by giving optimum temperature treatment depending on its method of synthesis, each phase illustrating distinct utilization as an oxygen ionic conductor, catalytic activity and ferroelectric materials. Three discrete phases mentioned as: at low temperature the stable phase of $Bi_2MoO_6$ is denoted by γ phase exhibiting fluorite-like structure, the intermediate phase is denoted by γ' which is stable till 600°C, in case of reversible transition, the γ and γ' phases show layered aurivillius type structure where perpetual arrangement of alternate $(Bi_2O_2)$ and $(MoO_4)$ sheets along a crystallographic axis. Recent studies by serval authors indicate that the transition of γ to γ' phase in $Bi_2MoO_6$ is basically associated with a ferroelectric to paraelectric type transition. γ' phase is generally orthorhombic in nature rather than that of tetragonal symmetry, in few studies, it is observed that at this temperature second harmonic generation signal generally disappears from $Bi_2MoO_6$ hence proving that γ' is non-polar in nature. Finally, a stable, non-reversible (γ") phase is obtained when Bismuth molybdate is heated above 670°C. The γ" phase has a monoclinic structure with space group of $P2_1/C$, γ" phase shows original structure due to presence of $MoO_4$ tetrahedra enclosing ribbons of Bi-O in place of $MoO_6$ octahedra which is generally revealed at lower temperature phases.[6-7] Physical properties of the material heavily depend upon the microstructure and doping, the effect of doping on different physical and chemical properties of the solid materials is well known and this response is widely employed in ferroelectrics to enhance their performance, so therefore by controlling these parameters, one can tailor the physical properties of the materials. Rietveld's X-ray powder structure refinement has been considered in the present work because no other method is equally capable of determining the crystal structure, lattice strain, crystallite size, and atomic parameters of nanocrystalline material consists of a large number of superimposed reflections.[8-9]

In this article, we determine the usefulness of new solid solution with the general composition of $Bi_2Mo_{1-x}W_xO_6$ by solid-state route method where x varies from 0.00 to 0.10. The phase, crystallite size and lattice parameters of $Bi_2Mo_{1-x}W_xO_6$ at room temperature is observed and deliberated on the basis of X-ray diffraction (XRD) with Rietveld

refinement,the change in phonon properties of composition due to appropriate doping, variation of electronic band gap is also investigating on basis of Uv-Vis spectroscopy, Raman scattering is carried out to gain evidence on structural changes occurring in the material.

## 2.Materials and methods

Stochiometrically taken $Bi_2O_3$, $MoO_3$ and $WO_3$ powders (All from Sigma Aldrich USA) were thoroughly grounded before microwave calcinations at 575°C for 5 hours. The range of tungsten doping in composition $Bi_2Mo_{1-x}W_xO_6$(BMoW) was from x=0.0 to 0.06 with stepsize 0.01, 0.08 and 0.10.Eurotherm 2416 controller was used for varying microwave input power monitored through a typical pre-set thermal profile during calcinations and sintering. Detailed preparation and initial x-ray diffractograms have been reported elsewhere[ ]. The refinement process of obtained x-ray diffraction data was carried using program Fullprof estimating P21c space group orthorhombic unit cell for prepared $Bi_2Mo_{1-x}W_xO_6$ (BMoW) compositions. The surface morphology of prepared pellets was investigated using field-emission SEM of make Carl Zeiss Merlin VP with 5kV acceleration voltage and 100KX magnification. Before recording images, pellet surfaces were coated using a thin layer of gold to avoid dipolar charging effects on surface. Microwave synthesis is the result of high frequency 2.45GHz microwaves coupled with each dipole in a material or atmosphere and generating local high temperature regions for calcinations/sintering due to dipolar switching at microwave frequencies. High-frequency treatment of starting dielectric oxide materials is For detecting transitions between vibrational energy states created for accommodating inelastic collision derived electron transfer, raman spectroscopy was deployed. These measurements were performed by exciting BMoW composition powders with He-Ne laser beam of 532nm wavelength. Scattered light was analyzed using an HR-800 Horiba JobinYvon, micro-Raman spectrophotometer having a spectral resolution of ~1 cm$^{-1}$ equipped with an edge filter(1800 lines/mm grating with CCD detector) for Rayleigh line rejection.Charge transfer mechanism in all BMoW compositions was investigated using impedance spectroscopy based on room temperature dielectric dispersion data and cyclic voltammetry. Dielectric dispersion (accuracy~0.08%) was recorded using high frequency LCR meter NF2376 (NF Corporation, Japan)in frequency range from 20 Hz to 1MHz at an oscillation level of 1V oscillation. Redox investigations were carried using cyclic voltammetry (C-V) performedon AutolabPotentiostatGalvanostat PGSTAT302N (Metrohm, Netherlands). For this electrochemical characterization,CV measurements were recorded in a three-electrodeset-up consisting of Ag/AgCl as the reference electrode,platinum wire as the counter electrode and glassy carbonelectrode (GCE) modified with two typical BMoW (x=0.0 and 0.10) compositionas the working electrode. The standard electrolyte usedin reaction was 5 mM of Potassium Ferro/Ferri cyanidein 0.1 M KCl. The CV scans were recorded from − 0.4 Vto 1 V with the scan rate of 0.50 mVs$^{-1}$. The *dc* conduction response of all BMoW compositions was recorded using conventional 2-probe setup connected with Keithley 6517B electrometer and Eurotherm 3216 controller based oven. The trigger voltages were varied from 1V to 10V and change in direct current was recorded with temperatures up to 300°C for determining low temperature defect/doping induced charge flow. *DC* conductivity was calculated using Ohm's law and each *dc* conduction behaviour was plotted in Arrhenius form for computing activation energy.UV-Vis spectrophotometer (Shimadzu Solid Spec-3700)was deployed for recording diffuse absorbance in an integrating sphere mode for estimating doping effects on optical energy bandgaps.Tauc plots were used for estimating optical energy bandgaps after confirming indirect band gap transition in BMoW materials.

## 3.Results and discussions

X-ray powder diffraction is one of the most powerful technique to obtain qualitative and microstructural information for polycrystalline materials. However, polymorphic forms of few multi-elemental ceramic matrices(sharing nearly equal lattice spacing with different crystal symmetries) are difficult to be analysed just by using as obtained diffraction data. For such cases, the quantitative study is not possible due to difficulties in the overlapping of the different Braggs reflections making traditional x-ray diffraction method unsatisfactory. To resolve this problem, a very well established Rietveld refinement method that is a simulation program to refine the theoretical line profile by least square fitting method, is used. Various profile-fitting parameters make it very convenient to conclude exact crystal symmetry or changes in parent structure besides helping in estimating strain induced microstructure. It is practiced to know the domain size, microstructure and strain values from the refined profile width parameters.Where $W_i$ is the weight parameter and $I_{io}$ and $I_{ic}$ is defined as observed and calculated intensity for diffraction angles $2\theta_i$. Other important parameters like profile residual ($R_p$) and weighted profile ($R_{wp}$) which is basically the measure of the degree of difference of calculated and observed values, is depicted below,[11-14]

$$\Delta = \sum_i W_i\{I_{io} - I_{ic}\}^2$$

$$R_p = \frac{\sum_i |I_{io} - I_{ic}|}{\sum_i I_{io}}$$

$$R_{wp} \left[\frac{\sum_i W_i(I_{io} - I_{ic})^2}{\sum_i W_i\{I_{io}\}^2}\right]^{1/2}$$

Fig.1 shows the Rietveld refined XRD patterns of polycrystalline $Bi_2Mo_{1-x}W_xO_6$(for x=0.0 and 0.06). The diffraction patternswere indexed assuming orthorhombicunit cell structure with space group *P2₁/c*.Excellent fitting can be seen through a close overlap between observed and calculated intensity values. This confirms the correctness of presumed orthorhombic unit cell structure and induces to explore exact structure of unit cell before and after doping. The profile matching indicates that unit cell remains unchanged post tungsten doping in place of molybdenum in octahedron cages. Refined derived atomic coordinates, listed in Table-1, were further used to obtain wycoff positions of each atom using Bilbao crystallographic server. All atomic coordinates indicate expected *8b* wycoff positions occupied by Bi, Mo/W and O ions.

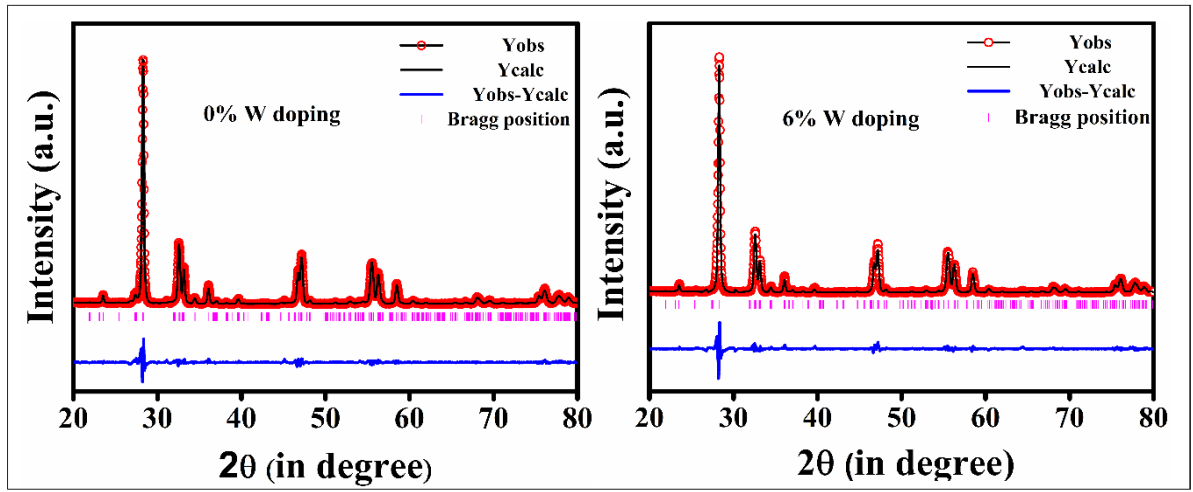

**Fig. 1. Rietveld refinement patterns of $Bi_2MoO_6$ and $Bi_2Mo_{0.94}W_{0.06}O_6$.**

**Table-1 Change in atomic positions on doping in $Bi_2Mo_{1-x}W_xO_6$ (with refined Wyckoff positions).**

| Atom coordinates (Wyckoff position) | $Bi_2MoO_6$ | $Bi_2Mo_{0.94}W_{0.06}O_6$ | $Bi_2Mo_{0.90}W_{0.10}O_6$ |
|---|---|---|---|
| **Bi(1) (8b)** | | | |
| X | 0.5242 | 0.5199 | 0.5204 |
| Y | 0.4208 | 0.4220 | 0.4245 |
| Z | 0.9945 | 0.9205 | 1.0413 |
| | | | |
| **Bi(2) (8b)** | | | |
| X | 0.4849 | 0.4816 | 0.4876 |
| Y | 0.0765 | 0.0786 | 0.0816 |
| Z | 0.9916 | 0.9167 | 1.0559 |
| | | | |
| **Mo(1)/W(1) (8b)** | | | |
| X | -0.0076 | 0.0080 | 0.0158 |
| Y | 0.2465 | 0.2459 | 0.2482 |
| Z | 0.0116 | 0.0707 | 0.0183 |
| | | | |
| **O(1) (8b)** | | | |
| X | 0.0196 | 0.0216 | 0.0266 |
| Y | 0.1446 | 0.1406 | 0.1291 |
| Z | 0.0901 | 0.0258 | 0.1327 |
| | | | |
| **O(2) (8b)** | | | |
| X | 0.2290 | 0.2986 | 0.3771 |
| Y | 1.009 | 0.9873 | 1.0486 |
| Z | 0.2659 | 0.2875 | 0.1796 |
| | | | |
| **O(3) (8b)** | | | |
| X | 0.2872 | 0.2811 | 0.1860 |

|          |   |        |        |        |
|----------|---|--------|--------|--------|
|          | Y | 0.5255 | 0.5026 | 0.5204 |
|          | Z | 0.3194 | 0.2150 | 0.3617 |
|          |   |        |        |        |
| O(4) (8b) |  |        |        |        |
|          | X | 0.7326 | 0.7501 | 0.6570 |
|          | Y | 0.2226 | 0.2260 | 0.2531 |
|          | Z | 0.2861 | 0.2263 | 0.4205 |
|          |   |        |        |        |
| O(5) (8b) |  |        |        |        |
|          | X | 0.2494 | 0.2060 | 0.1847 |
|          | Y | 0.3242 | 0.2769 | 0.2910 |
|          | Z | 0.5670 | 0.2942 | 0.3610 |
|          |   |        |        |        |
| O(6) (8b) |  |        |        |        |
|          | X | 0.6187 | 0.6311 | 0.6368 |
|          | Y | 0.3834 | 0.3335 | 0.3522 |
|          | Z | 0.5453 | 0.5221 | 0.6983 |

Absence of *4a* wycoff positions in all prepared BMoW materials confirm the formation of first layer Aurvillius phase formation on doping. The refined unit cell parameters and profile reliability parameters for each BMoW composition are listed in Table-2. A marginal deviation in the values of statistical parameters $R_p$, $R_{wp}$ and $R_{exp}$ for all doped BMoW compositions compared to undopedBMo composition establishes the following, *a)* absence of any undesired polluting phase throughout *even* doping range and *b)* unchanged orthorhombic unit cell in all BMoW compositions[15]. Careful comparative analysis of experimentally obtained lattice parameters with those refined indicates prominent *a-b* plane octahedral rotations on tungsten doping than *a-c* plane tilting of the same, Table-3. This induces further in-depth study of tungstendoping caused changes in bond angles of BMo unit cells, Table-4, and leads for the designing of unit cell structure as shown in Fig.2. Tungsten [Xe: $5d^4$, $6s^2$] intrinsically possess lesser electron-electron repulsions than Molybdenum [Kr: $4d^5$, $5s^1$] having more diffused charge distribution. This renders a vacant 5$^{th}$ $d$-orbital ($d_{x2-y2}$) in tungsten compared to molybdenum generating more free rattling space in *a-b* plane. Hence, Mo/W bonds with selective oxygen's (e.g. O1, O4, O5 and O6) show significant rotations than the others (O2 and O3). This is indicated in Fig.2, where octahedrons $MoO_6$/$WO_6$ appear to rotate significantly along –*c-b* plane clockwise and in complementary manner towards each other (O1, O4 and O5 can be prominently seen to re-aligned by octahedral tilting) [R.G.Teller, ActaCryst paper, 1984]. Therefore, tungsten doping in BMo matrix induces octahedral tilting to keep oxygen ions away from open surfaces designing these surfaces typically oxygen free for any surface driven effect.

**Table-2 Unit cell parameters and reliability factors for Rietveld refined doped $Bi_2Mo_{1-x}W_xO_6$ crystal structure**

| Parameters | a (Å) | b (Å) | c (Å) | $R_p$ (%) | $R_{wp}$(%) | $R_{exp}$ |
|------------|-------|-------|-------|-----------|-------------|-----------|
| x=0.00     | 5.485 | 16.209 | 5.507 | 12.4 | 15.8 | 8.03 |
| x=0.02     | 5.487 | 16.213 | 5.508 | 15.3 | 18.1 | 9.9  |
| x=0.04     | 5.485 | 16.216 | 5.5066 | 14.3 | 19.1 | 8.64 |
| x=0.06     | 5.484 | 16.219 | 5.507 | 10.72 | 13.2 | 4.68 |
| **x=0.08** | 5.484 | 16.217 | 5.508 | 15.2 | 17.7 | 9.77 |
| **x=0.10** | 5.484 | 16.217 | 5.509 | 13.9 | 17 | 8.7 |

**Table-3 Unit cell parameter variation in $Bi_2Mo_{0.94}W_{0.06}O_6$ before and after refinement.**

| Parameters | Before Rietveld Refinement | After Rietveld Refinement |
|---|---|---|
| a (Å) | 5.406 | 5.4839 |
| b (Å) | 16.509 | 16.2177 |
| c (Å) | 5.503 | 5.5056 |

**Table-4 Change in bond angles with doping in $Bi_2Mo_{1-x}W_xO_6$ (all values are in degrees).**

| Atom | x=0.0 | x=0.06 |
|---|---|---|
| Bi2-O1-Mo1/W1 | 40.7035 | 58.7880 |
| O6-Mo1-O4 | 102.574 | 69.2969 |
| O1-Bi1-Mo1 | 49.7476 | 16.5515 |
| Bi1-Mo1-O5 | 105.2259 | 80.0790 |
| W1/Mo1-Bi1-O6 | 150.9642 | 32.1484 |
| W1/Mo1-Bi2-O2 | 145.9537 | 142.9503 |
| W1/Mo1-Bi2-O5 | 33.3456 | 42.8488 |

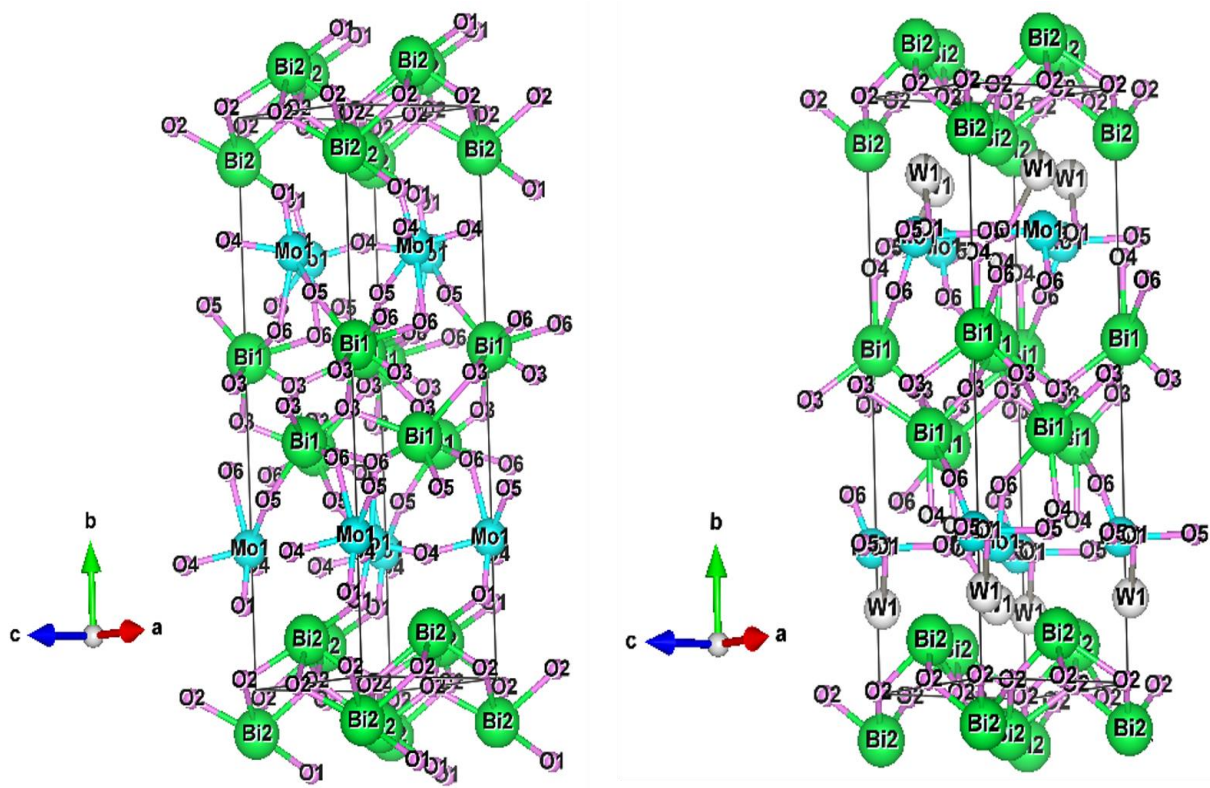

**Fig.2. Crystal structure of $Bi_2MoO_6$ (x=0.0) and $Bi_2Mo_{0.94}W_{0.06}O_6$ (x=0.06)**

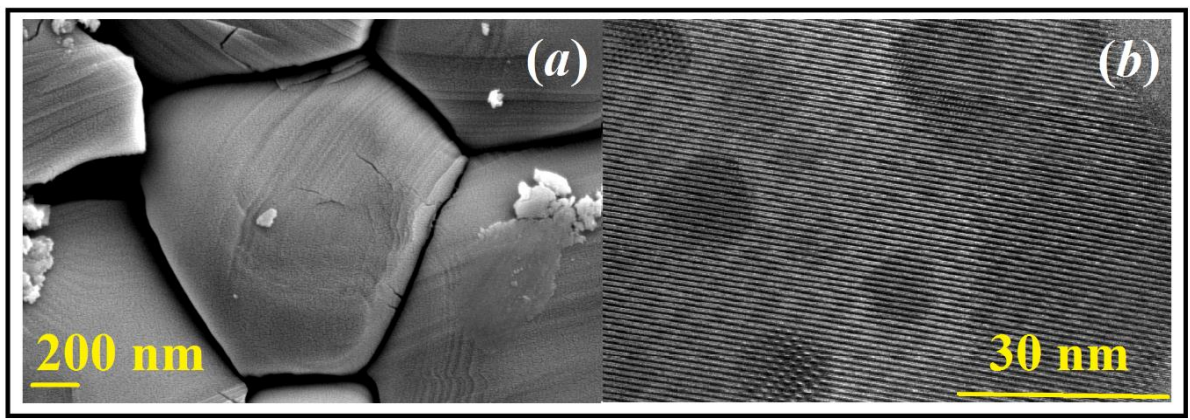

**Fig.3 Grain distribution and nanoparticle formation in pure Bi$_2$MoO$_6$**

Hydraulically compacted and microwave sintered (at 595°C for 2hrs) pellet surface of Bi$_2$MoO$_6$ composition has been shown in Fig.3(a) using field-emission SEM micrographs. An interesting hexagonal grain formation with *alternately arranged* equal dimensions. Such a continuous grain formation occurs in powders with schematically distributed pores usually derived in microwave synthesis [ ]. All grains have edge-to-edge dimension of 1.8µm, that is very well within standard range (1-2.5µm) of grain size in bismuth layer materials due to bismuth oxide loss from grain connecting surfaces during sintering to coalesce grains in better manner [ ]. Average grain size is estimated to be 3μm using ImageJ software that presumably incorporates width of darker void boundaries into estimation. Fig.3(b) depicts the TEM image of crushed pellet, hexagonal nano particles (size ~15-20nm) grown on regular atomic network can be seen. Such jutted out nano particles are overall dispersed on micro sized grains possibly due to trapping of microwaves between molybdenum-sites of gross network acting as *susceptors*. Microwaves trapped between these sites heat the region excessively and major bismuth oxide region swallows towards surface taking all interior planes of Mo/W-oxides together.

Apart from XRD, Raman scattering investigations help crucially in determining, *a)* phase purity, *b)* effect of dopant on modifying local vibrational behaviour of metal-oxygen bonds

(like stretching/bending/wagging) and *c*) finally electronic shell deformation effect on intrinsic electric dipole moment. Raman spectra of typically chosen $Bi_2Mo_{1-x}W_xO_6$ compositions is shown in Fig.4, all expected raman bands are observed. Intensity of the raman bands change optimally with tungsten doping in $Bi_2Mo_{1-x}W_xO_6$ matrix. Raman spectra is a close fingerprint of polymorphic structural changes that occur in $Bi_2MoO_6$ and $Bi_2WO_6$ materials mostly due to the rigid rotations of $MoO_6/WO_6$ octahedrons [Maczka PRB (2008), Maczka J.Phys.:Condensed Matter (2010)]. Raman bands observed above 600cm$^{-1}$ belong to the Mo-O stretching and below 400cm$^{-1}$ to bending, wagging and external translational motion of Bi- and Mo-ions. Intense raman bands observed around 793cm$^{-1}$ and 850cm$^{-1}$ in present work are corresponding to symmetric ($A_{1g}$ mode) and asymmetric ($A_{2u}+E_u$ mode) stretching of the $MoO_6/WO_6$ octahedrons. Both of these modes emerge as non-centrosymmetric and polar due to intrinsic structural distortions in MoO6 octahedrons partly compensated by larger electronic cloud of tungsten (W). Additional bands in the range 690-720cm$^{-1}$ are due to the asymmetric stretching mode of MoO$_6$octahedravia motion of equatorial oxygen ions connecting these octahedronswith Bi$_2$O$_2$layers. The bending modes for bismuth oxide polyhedral connected with MoO$_6$ octahedrons are observed from 180-500cm$^{-1}$. Raman active doubly degenerate ($E_g$) modes due to rocking of the octahedronsare observedfrom 270cm$^{-1}$to 360cm$^{-1}$. Optimum 6% tungsten doping in Bi$_2$Mo$_{1-x}$W$_x$O$_6$matrixat ambient pressure (0.0001GPa) and temperature (25ºC) generates non-centrosymmetric polar orthorhombic structure identical to the Bi$_2$MoO$_6$matrix prepared under extreme high pressure (~8.3GPa) using diamond anvil setup. Tungsten doping beyond 6% is observed to saturate polar character of Bi$_2$Mo$_{1-x}$W$_x$O$_6$ unit cell and larger electronic cloud of tungsten starts shielding intrinsic dipoles to result into less polar or more symmetric unit cell possibly in tetragonal space group I4/mmm [R.L.WithersJ.Solid State Chem, 1991].

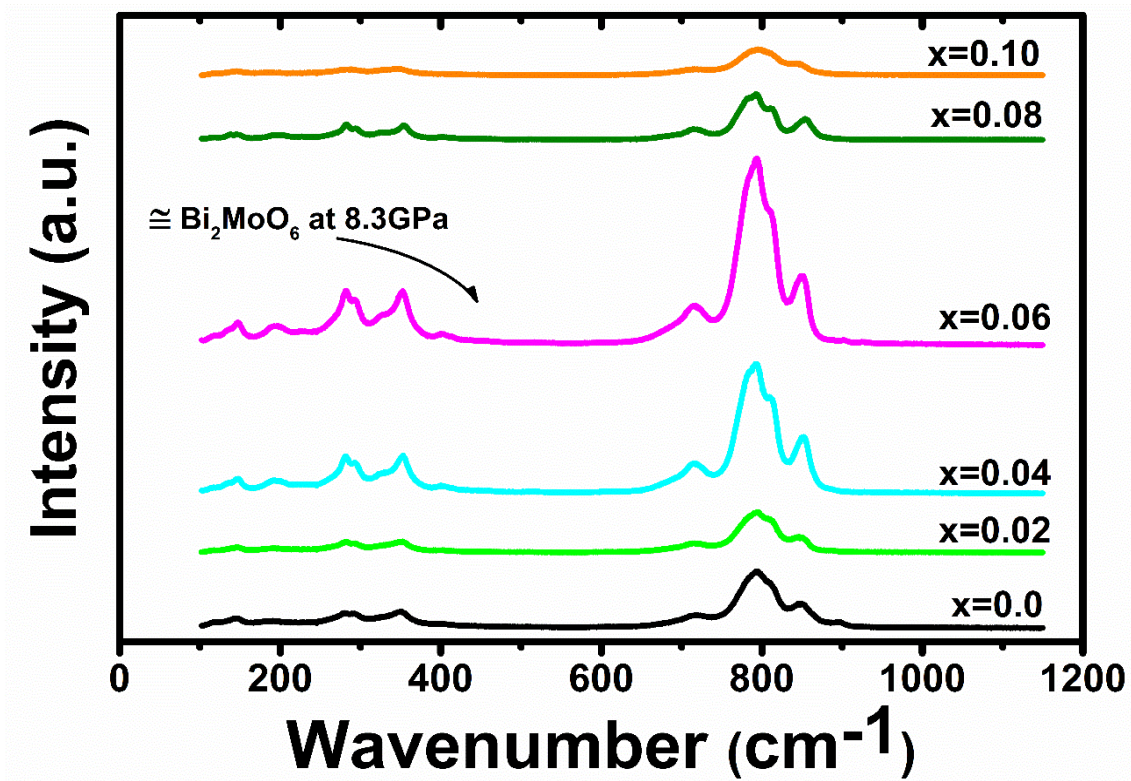

**Fig.4 Raman spectra of $Bi_2Mo_{1-x}W_xO_6$ compositions recorded at 25ºC and 0.0001GPa**

Complex Impedance spectroscopy (CIS) isan effective tool that can resolveelectronic conduction deriving from all possible sources like grain, grain boundaries and grain-electrode interface in applied electric input frequency domain. Nyquist plots (Z'-Z'') areused in determining conduction from major grain resistance part of the material though not sensitive towards conduction corresponding to smaller values of resistancedue to grain boundary and electrode-material interface. This low resistance conduction mechanism is explained by complex modulus plot (M'-M''). Fig.5shows the room temperature complex impedance plot for doped Bi$_2$Mo$_{1-x}$W$_x$O$_6$materials investigated from 100Hz to 1MHz. Bulk grain resistance values for tungsten-doped Bi$_2$Mo$_{1-x}$W$_x$O$_6$ samples are sufficiently lower than undoped Bi$_2$MoO$_6$ material. Tungsten doping~2% was able to curtail lossy behaviour of pure Bi$_2$MoO$_6$material that shows non-Debye tail at low frequencies. This is due to comparatively large resistive-capacitive grain boundary exhibiting multiple relaxation times. All tungsten-doped Bi$_2$Mo$_{1-x}$W$_x$O$_6$samples show nearly an ideal Debye behaviour with centres of all semicircles well above Z'-axis and emerging 2nd semicircle optimally around 4-6% tungsten doping. For samples doped beyond 6%, the 2nd semicircle emerges again and becomes wider. Fig.6 shows the frequency dispersion of Z' and Z" for $Bi_2Mo_{1-x}W_xO_6$ materials. Real impedance Z' for all tungsten doped samples decreases to nearly 10% of its value for undopedBMo sample. Overall, it is minimum for 2% tungsten doping and maximum for undoped $Bi_2MoO_6$. The distribution of Z'-curves before $10^3$Hz is in order of space charge magnitude in $Bi_2Mo_{1-x}W_xO_6$ materials that saturates quickly beyond $10^5$Hz irrespective of tungsten concentration. In accordance with Z'-behaviour, Z"-peaks shift regularly towards low frequencies till 8% tungsten doping in $Bi_2Mo_{1-x}W_xO_6$ materials indicating maximum loss of energy at resonance due to low frequency responsive space charge accumulating at grain boundaries and inhibiting charge transfer between grains due to high capacitive reactance ($X_c$). Minimum height of Z"-peak for 2% tungsten doped sample is an indicative of minimum bulk grain resistance along with resistive grain boundary. An increase in peak broadening on increasing tungsten content from 2 to 8% is an indicative of doping caused multiple relaxation that decreases as tungsten concentration is 10%. Modulus spectroscopy distinguishes grain and grain boundary capacitive contribution by suppressing electrode effect. Fig.7 shows the conclusive tungsten doping control on high frequency responsive capacitive grain boundary and low frequency responsive space charge boundaries between material and electrodes in $Bi_2Mo_{1-x}W_xO_6$ materials. Highly discontinuous (in high frequency range) pair of two semicircles for undoped $Bi_2MoO_6$ material shows the elimination of highly capacitive porous grain boundary formed due to microwave synthesis. Initial introduction of tungsten by 1-2% is able to dissolve this capacitive grain boundary via high solid solubility feature. As a result, only low frequency responsive space charge boundaries could survive post synthesis and short after applying high frequency fields.

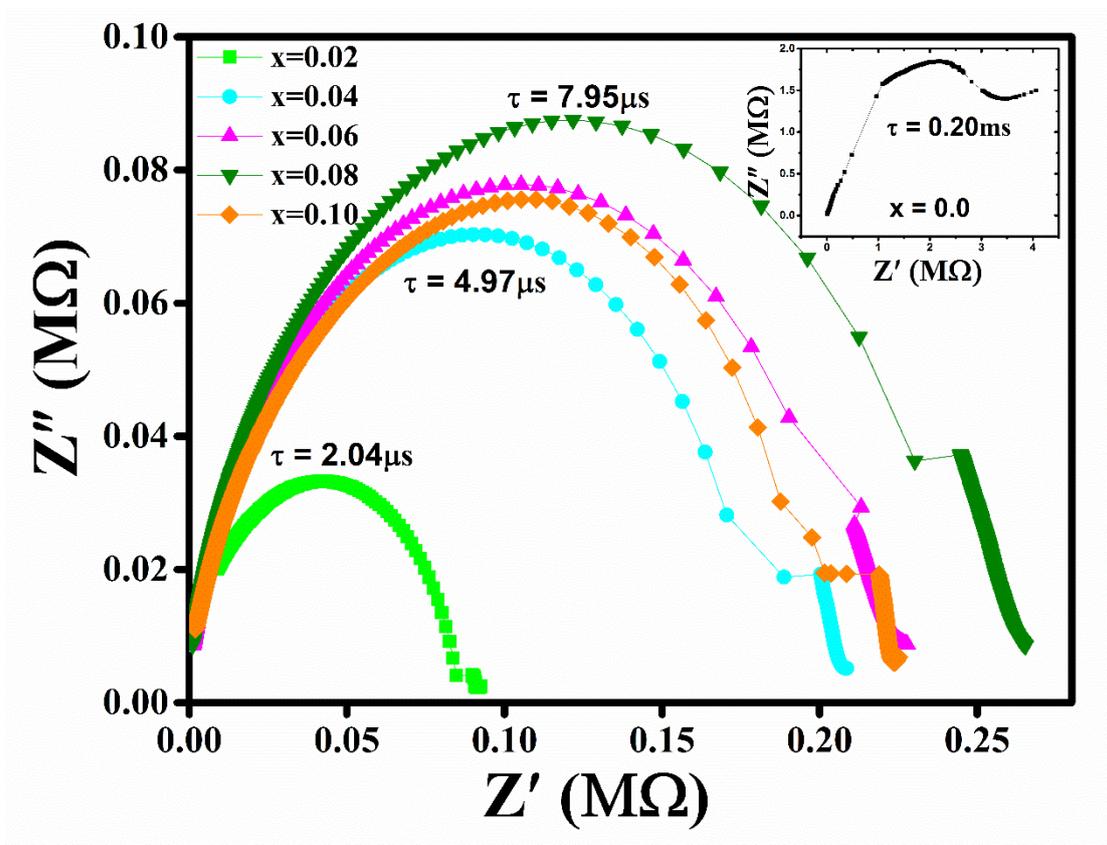

**Fig.5 Nyquistplots of $Bi_2Mo_{1-x}W_xO_6$ materials**

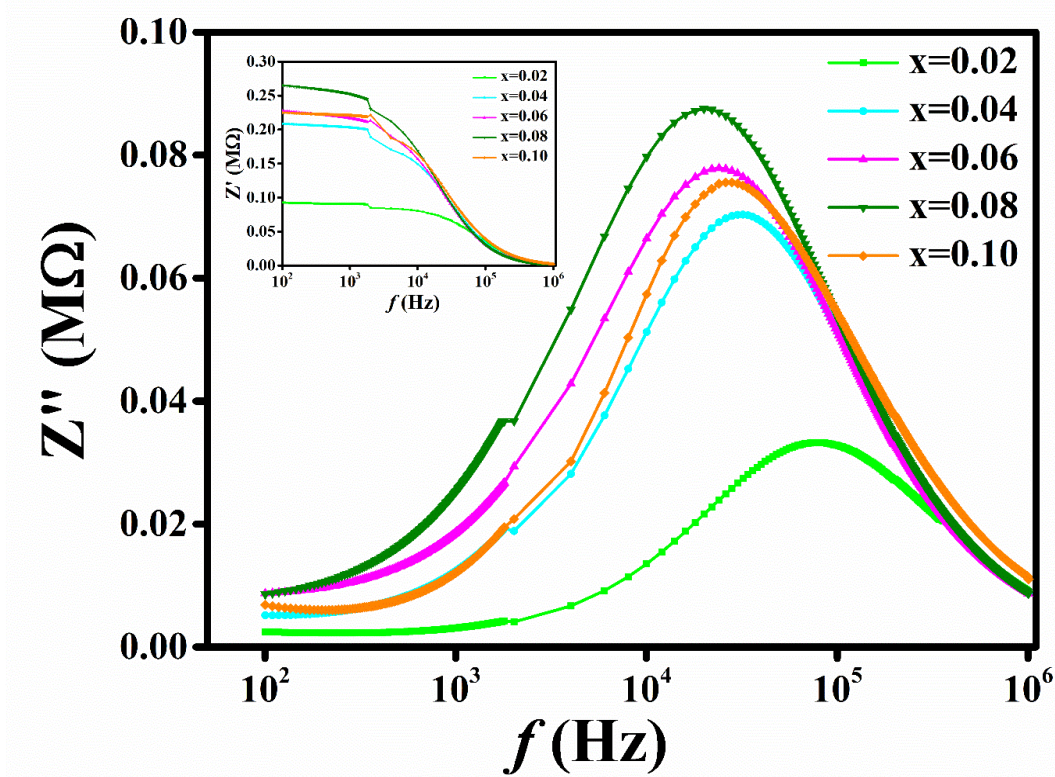

**Fig.6 Frequency dispersion in real and imaginary impedance**

Therefore, doping range 4-8% shows minimum capacitance contribution (via intercept of semicircles on M'-axis) in $Bi_2Mo_{1-x}W_xO_6$ materials compared to undoped, 2 and 10% doped ones.

Long-range mobility is the hopping mechanism of electronic conduction between grains and short-range mobility is the localized conduction through diffused and conducting grain boundaries. Modulus spectroscopy is very effective tool in distinguishing these two mechanisms. Fig.8 represents the frequency dependence of real and imaginary part of electric modulus (M' and M") for tungsten-doped BMO materials. Real modulus (M'),for each tungsten doping step, increases with frequency and saturates to a new constant value ($M_\infty$)at frequencies beyond 1MHz. The distribution of $M_\infty$-valueis in correlation with effect of tungsten doping in BMO materials discussed in earlier sections, i.e., minimum for 2% doped sample and maximum for undoped and 8% doped sample. This dispersion of real modulus (M') is also an indicative of dominant grain conduction over grain boundaries due to short-range mobility of electrons/holes. Imaginary modulus (M") for each doped BMoW material exhibits an asymmetric peak depicting two distinct regions of short-range mobility (localized conduction) and long-range (hopping conduction) mobility of charge carriers. Below peak, it is the range in which charge carriers are mobile over long distances and beyond peak value, these are confined in atomic potential wells to move for short distances. Undoped BMO material exhibits long-range mobility due to microwave synthesis derived schematic pore distribution, Fig.S1. Tungsten doping ~ 1-2% is capable to diffuse the pore formation and arrange short-range mobility for electrons and holes. Therefore, non-Debye behaviour (hopping behaviour) of undopedBMois effectively suppressed by tungsten doping in low range from 1-2%.

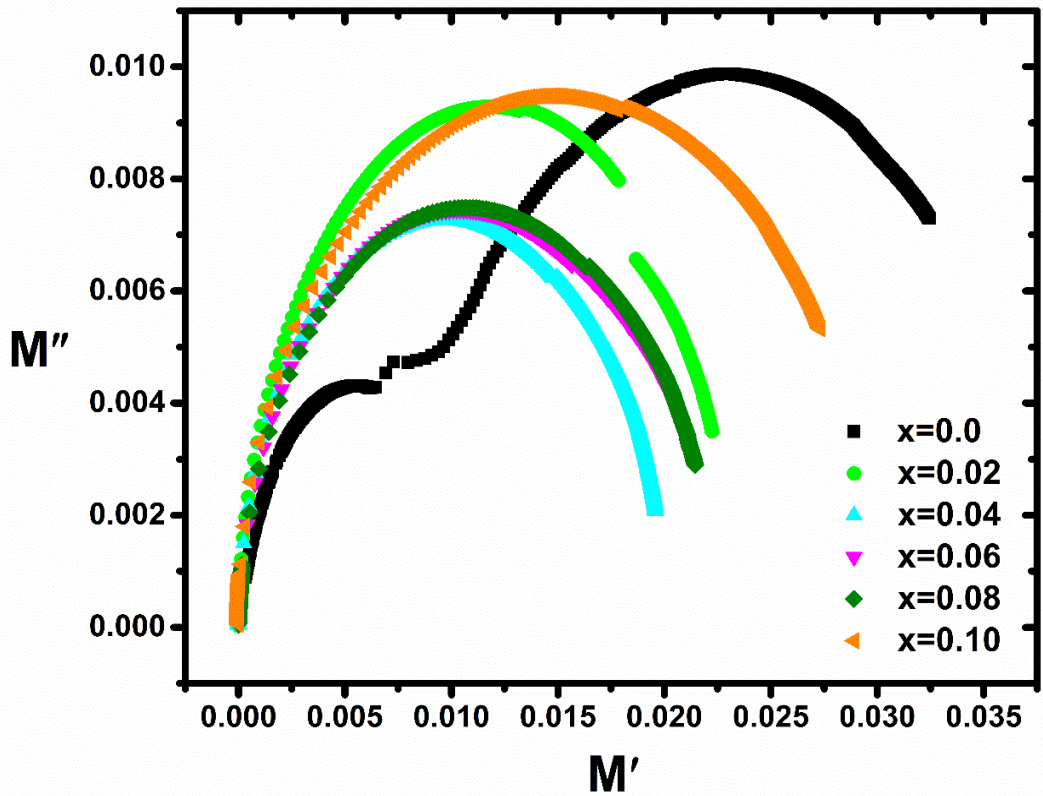

**Fig.7 Frequency dependence of real (M') and imaginary part (M'') of electric modulus**

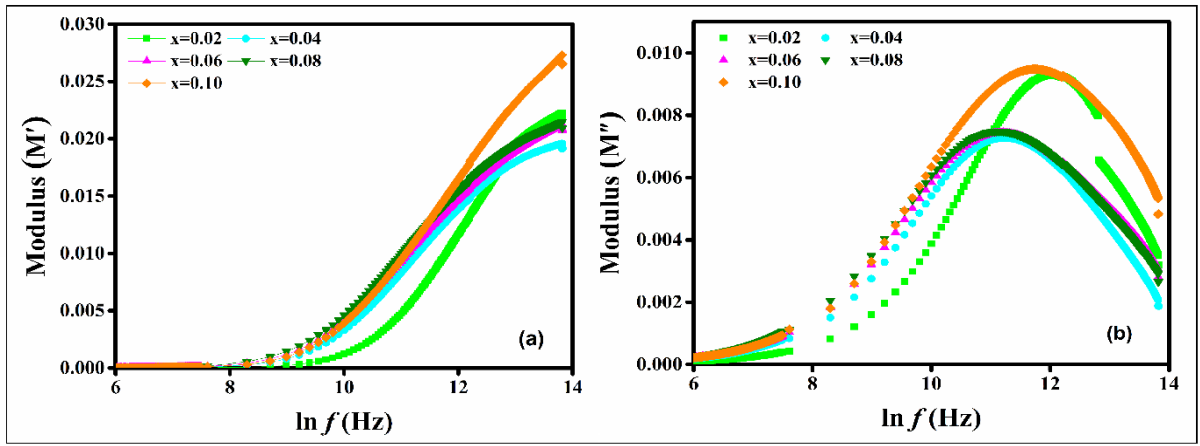

**Fig.8 Frequency dependence of real (M') and imaginary part (M'') of electric modulus**

Non-capacitive Nernstian charge storage mechanism is clearly seen in BMoW electrodes, Fig.9. Typical BMoW compositions, undopedBMoand 10% W-doped BMo are modified using glassy carbon electrode (GCE) by dissolving 5 $m$MFe(CN)$_6^{3-/4-}$ redox couplein 0.1 M KCl. The comparison of peak currents ratio along with peak potential difference confirms an increase in reversibility of tungsten doped BMo faradaic system [BMo(I$_o$/I$_r$ = 0.97, $\Delta$E$_p$=0.1510 V);BMoW(I$_o$/I$_r$ =1.05, $\Delta$E$_p$= 0.1806 V)]. Such an increase in peak potential difference is due to decreased rate constants ($k_{red}$/$k_{ox}$) depending upon Gibbs free energy change ($\Delta G$) and excess redox potential ($V$):

$$k_{red} = Z \exp\frac{-\Delta G_{red}^{V=0}}{k_B T} \exp\frac{-\alpha FV}{k_B T}$$

$$k_{ox} = Z \exp\frac{-\Delta G_{ox}^{V=0}}{k_B T} \exp\frac{-(1-\alpha)FV}{k_B T}$$

where other symbols have their usual meaning (Ref). Decrease in peak current on introducing tungsten in Bi$_2$MoO$_6$ composition is due to reduction in concentration of redox molecules though generation of unsaturated charge centres for higher valence states (+3/+4) is seen.This is estimated through minor increase in anodic current around 0.75V. Charge transfer resistance ($R_{ct}$) of an electrode is known to decrease on increasing redox molecule concentration; the same is confirmed by comparing radius of Nyquist plot for undopedBMo sample with that of BMoW$_{0.10}$ sample as shown in Fig.7. Lower radius of plot for undopedBMo sample compared to BMoW$_{0.10}$ sample confirms low charge transfer resistance ($R_{ct}$) hence large redox current. Tungsten modified GCE/BMoW$_{0.10}$ sample compared to GCE/undopedBMo is two-electron transfer compared to single electron transfer in BMo. The ratio of the anodic to cathodic peak currents is lower in BMo electrode due to nearly all molybdenum ions (reducing) getting consumed in the subsequent chemical reaction, resulting in fewer atoms to oxidize during anodic scan. On the other hand, tungsten (in BMoW$_{0.10}$) reduces by accepting charge from Fe2+, Fe3+ and Fe4+ ions due to vacant *d*-orbital structure and oxidizes accordingly for more atomic sites during anodic scan.

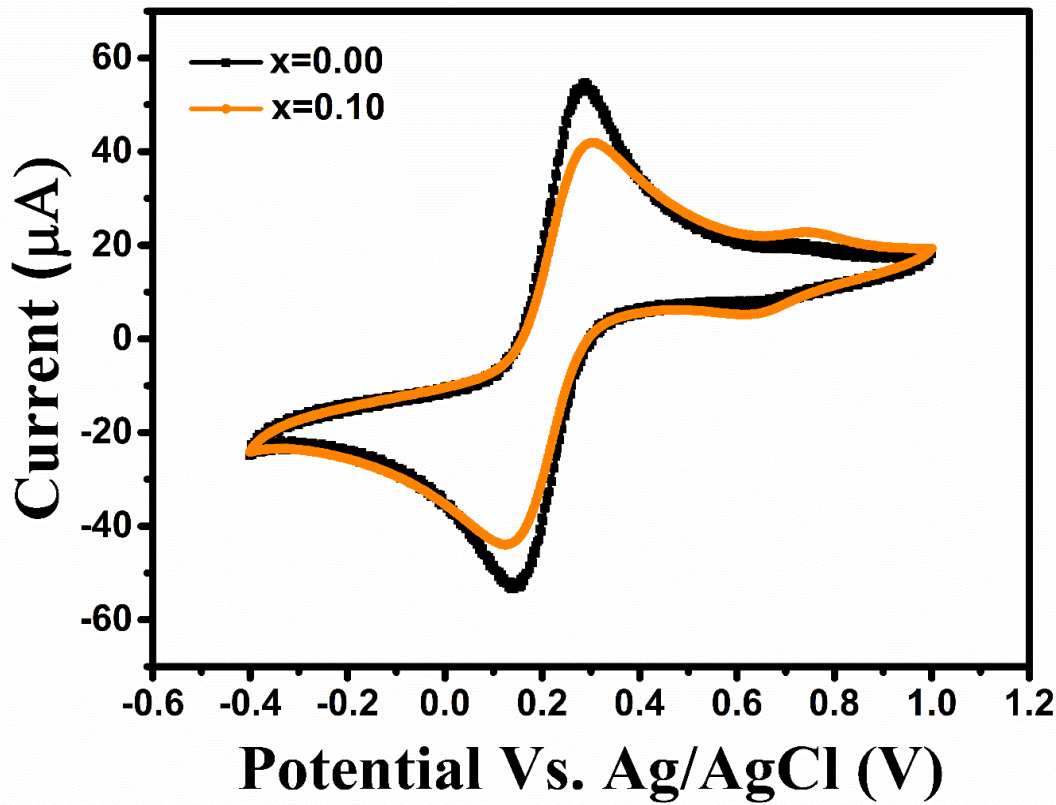

Fig.9 Multiple redox regions in tungsten doped Bi$_2$Mo$_{1-x}$W$_x$O$_6$ material

Impedance spectroscopic data as shown in Fig.5 confirms inherent oxide ion conduction dominance in Bi$_2$MoO$_6$ material (Ref-Ref-Ref)). Low frequency spike emerging prominently in case of undoped Bi$_2$MoO$_6$ sample is an indicative of oxygen ion migration in lattice. Tungsten doping curtails the length of this spike significantly and converts into a low radius semi-circular arc. This is very interesting to notice that generation of freely migrating oxygen ions in Bi$^{3+}$-Mo$^{6+}$ networked lattice get limited to outer surfaces only in Bi$^{3+}$-Mo$^{6+}$-W$^{6+}$ networking. This restructuring of spike into semi-circular arc to drive dominating electronic conduction over ionic conduction is at its best for 2% tungsten doping. This provides a hint of pre-existing 2% bismuth vacancies in $Bi_2MoO_6$ material those are being saturated on tungsten doping. Doping of tungsten beyond 2% keeps the semi-circular form of spike maintained though with slightly increased radius under reduced height. This is an indicative of resistive-capacitive grain boundary formation away from diffusion derived Warburg resistance on electrode-grain boundary also concluded in discussion on fig.7. A few authors interpret appearance of dominating Warburg resistance in $Bi_2MoO_6$ material due to anisotropic thermal response of Mo-O bonds in low temperature rangefor increasing conducting [Ref-R.MuruganPhysica B]. DC conductivity values as listed in table-3 support the interpretation on Fig.5 and further supplement the understating of electronic conduction building over oxygen ion conduction in $Bi_2Mo_{1-x}W_xO_6$ materials.Multiple activation zones can be seen in dc conductivity response of $Bi_2Mo_{1-x}W_xO_6$ materials determining role of tungsten doping in easing electron transfer, fig.10. Nearly single linear response is obtained for 6-8% tungsten doping; this confirms maximized ease in electron transfer in wide thermal range from 25°C to 300°C. However, distinctly seen two and three linear regions are recorded for other low doped $Bi_2Mo_{1-x}W_xO_6$ materials. These regions exhibit emergence of thermal actuation beyond 120°C and possessing inflexions beyond 180°C. The activation energies are calculated from Arrhenius plots and are reported corresponding to inflexion point in Table-3. The reason of different activation regimes has been interpreted earlier in terms of blocking oxygen ion transport in $Bi_2Mo_{1-x}W_xO_6$lattice. High activation energy is due to the high polarizability of $Bi^{3+}$ cationsdue to lone pair of electronsthose occupy free space associated with the $O^{2-}$ anion and hinder its ($O^{2-}$) movement in its sub-lattice under *dc/ac* fields.

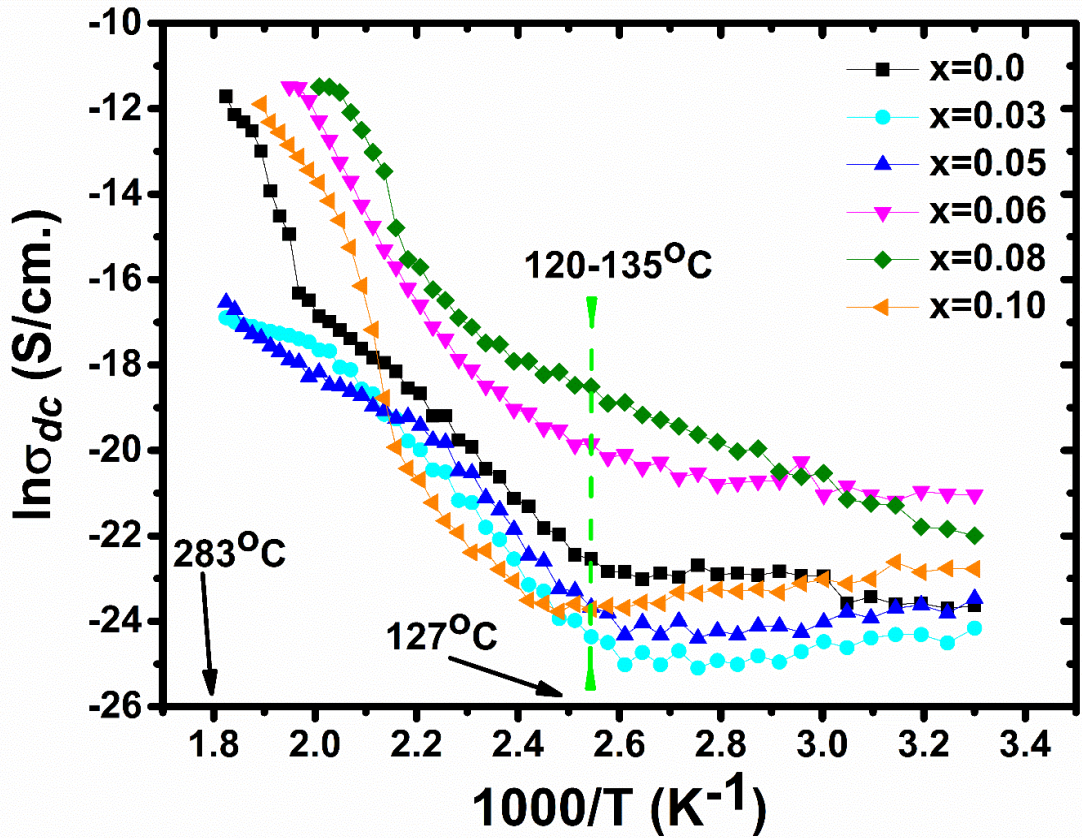

**Fig.10** Multiple thermal activation zones in tungsten doped $Bi_2MoO_6$ materials

Table. 3 shows the variation of the electrical and optical energy band gaps for $Bi_2Mo_{1-x}W_xO_6$ materials

| Tungsten doping (%) | $\sigma_{dc}$ at 25°C ($\times 10^{-12}$ S/cm.) | $\sigma_{dc}$ at 50°C ($\times 10^{-12}$ S/cm.) | $\sigma_{dc}$ at 250°C ($\times 10^{-6}$ S/cm.) | Electrical energy band gap (eV) | |
|---|---|---|---|---|---|
| | | | | 135-180°C | 180-280°C |
| 0 | 54.40 | 66.76 | 0.89 | 0.95 | 2.32 |
| 3 | 32.05 | 25.56 | 0.033 | 1.12 | 0.27 |
| 5 | 64.09 | 40.69 | 0.023 | 1.10 | 0.61 |
| 6 | 726.14 | 731.08 | 10.23 | 0.66 | 1.89 |
| 8 | 279.66 | 594.93 | 10.30 | 0.44 | 2.27 |
| 10 | 128.33 | 101.49 | 4.48 | 1.13 | 1.43 |

Electronic states of low-doped $Bi_2Mo_{1-x}W_xO_6$ materialsare characterized by extremely useful *uv*-diffuse reflectance spectroscopy in an integrated sphere mode. Non-parallel response of all absorption curves is an evidence of existing band to defect energy states transition in addition to band-to-band transition. Initial blue shift in absorption edge on increase in tungsten doping until 6% is because of merging defect energy states in conduction band edge rendering a larger optical energy gap. Doping beyond 6% generates additional energy states due to increase in polarizability and scattering effects. Optical bandgap energies are calculated using Kubelka-Munk function: $(\alpha h\nu)^n = A(h\nu - E)$, where *α*, *h*, *ν, E* and *A*are absorption coefficient, planck's constant, frequency of electromagnetic radiation, band gap energy and a medium dependant constant respectively. Here, *n* determines the direct (*n*=2) or indirect (*n*=1/2) band transition characteristic of the materials. $Bi_2Mo_{1-x}W_xO_6$ materials show direct band gaptransition based on linear response between (α*hν*)² and *hν* (*Tauc* plots). All energy gaps are in the range 2.48-2.75eV, such a 10% variation is very well within acceptable range according to tungsten doping concentration, Table-3. Also, the values of absorption edge onset and optical energy band gap for current microwave processed $Bi_2Mo_{1-x}W_xO_6$ materialshave close resemblance with nanocrystalline $Bi_2MoO_6$ materials prepared by solvothermal route [Ref-Actamaterialia]. Minor transition peak around 220nm is between additional energy states created for bismuth lone pair electrons that disappears completely at optimized tungsten doping level of 6% [Ref.].

**Table-4Variation of optical energy band gap tungsten-doped $Bi_2Mo_{1-x}W_xO_6$ materials**

| Tungsten Doping (%) | Onset of absorption edge (nm) | Optical Energy Band gap (eV) |
|---|---|---|
| 0 | 494.7 | 2.485 |
| 2 | 494.5 | 2.516 |
| 4 | 493.8 | 2.558 |
| 6 | 487.4 | 2.563 |
| 8 | 492.5 | 2.747 |
| 10 | 494.5 | 2.546 |

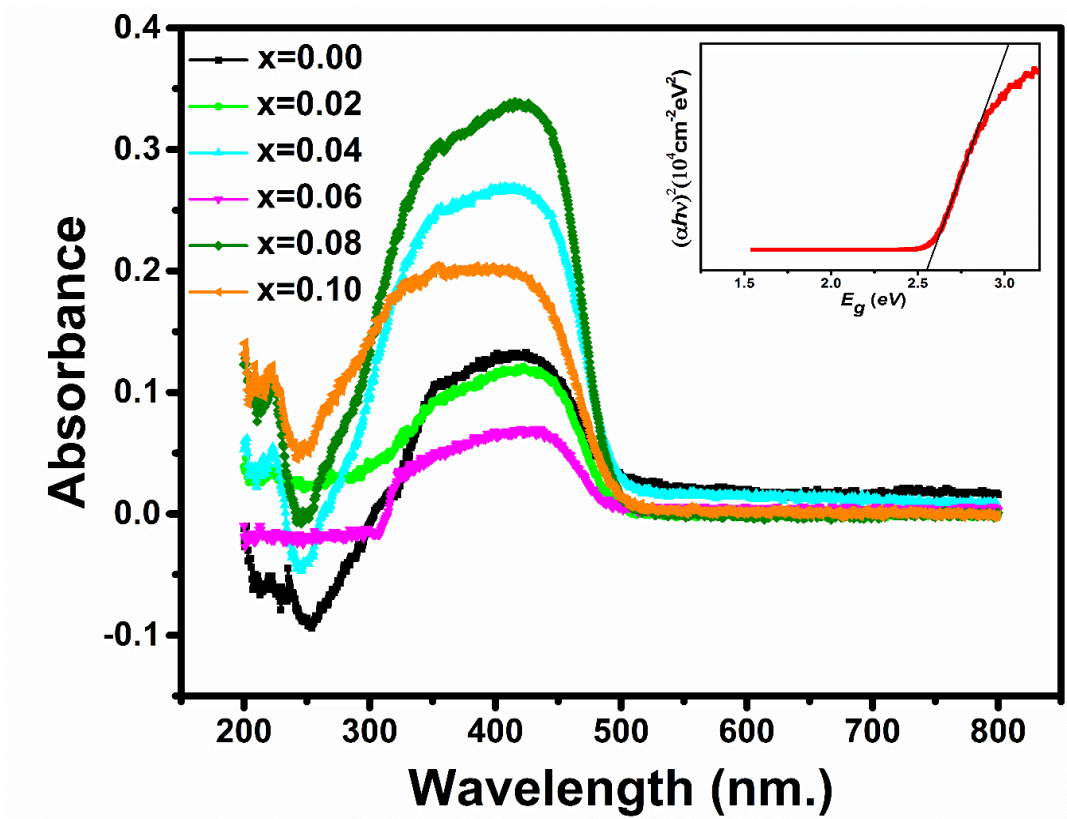

**Fig.11** Uv-vis diffuse absorption spectra of $Bi_2Mo_{1-x}W_xO_6$ (inset depicts typicalbandgap calculation using Tauc method for x=0.04 sample)

## 4. Conclusions

The polycrystalline Bismuth molybdates are prepared from conventional solid-state route method and its Rietveld refinement reveals the orthorhombic crystal structure with space group $P2_1/C$. A gradual increase in the band gap of BMO is observed with the introduction of tungsten in the system and this increase is also suggested by the increased activation energy.

Raman data suggest the structural changes in BMO is basically due to the distortion of $MoO_6$ octahedra. All this study suggest that BMO can be effectively used as a photocatalyst. The polycrystalline Bismuth molybdates is prepared from conventional solid-state route method and its Rietveld refinement revels the orthorhombic crystal structure with space group $P2_1/C$. An increase in *dc* activation energy is observed with incorporation of $W^{6+}$ in the system and give rise to insulating behaviour in the material. Non-Debye type behaviour with distributed relaxation time period is observed from complex impedance plot. Gradual increase in the band gap of BMO is observed with introduction of tungsten in the system and this increase is also suggested by the increasing activation energy. Raman data suggest the structural changes in BMO is basically due to distortion of $MoO_6$ octahedra. All this study suggest that BMO can be effectively used in sensor industry.

## Acknowledgments

Authors greatly acknowledge Dr. Vasant Sathe from UGC-DAE Indore for permitting us to record Raman data and Dr.Mukesh Ranjan with Mr. Suraj K. P from FCIPT-IPR Gandhi Nagar for providing Fe-SEM images. One of the authors Anurag Pritam expresses gratitude towards the Shiv Nadar Foundation for providing research fellowship.